\documentclass[apjl]{emulateapj}
\usepackage{apjfonts,amssymb,amsbsy}
\usepackage{graphicx}
\def\apj{ApJ}
\def\apjs{ApJS}

\def\apjl{ApJL}

\def\aap{A\&A}
\def\mnras{MNRAS}
\def\prl{Phys. Rev. Lett.}
\def\pre{Phys. Rev. E}

\def\lsim{~\raise0.3ex\hbox{$<$}\kern-0.75em{\lower0.65ex\hbox{$\sim$}}~}
\def\gsim{~\raise0.3ex\hbox{$>$}\kern-0.75em{\lower0.65ex\hbox{$\sim$}}~}
\journalinfo{The Astrophysical Journal Letters, 727:L20 (5pp), 2011 January 20} 
\submitted{Received 2010 July 16; accepted 2010 December 10; published 2010 December 29} 
\shorttitle{DENSITY PDF IN STAR-FORMING CLOUDS}
\shortauthors{KRITSUK, NORMAN \& WAGNER}

\begin{document}
\title{On the Density Distribution in Star-forming Interstellar Clouds}

\author{Alexei G. Kritsuk,$^1$ Michael L. Norman,$^{1,2}$ and Rick Wagner$^2$}
\affil{$^1$Physics Department and Center for Astrophysics \& Space Sciences,
University of California, San Diego, 9500 Gilman Drive, La Jolla, CA 92093-0424, USA\\
$^2$San Diego Supercomputer Center,
University of California, San Diego,  10100 Hopkins Drive, La Jolla, CA 92093-0505, USA}

\begin{abstract}
We use deep adaptive mesh refinement simulations of isothermal self-gravitating supersonic turbulence
to study the imprints of gravity on the mass density distribution in molecular clouds.
The simulations show that the density distribution in self-gravitating clouds develops an
{\em extended} power-law tail at high densities on top of the usual lognormal. We associate the
origin of the tail with self-similar collapse solutions and predict the power index values 
in the range from $-7/4$ to $-3/2$ that agree with both simulations and observations of 
star-forming molecular clouds.
\end{abstract}


\section{Introduction}
The probability density function (PDF) of the mass density in non-self-gravitating isothermal
supersonic turbulence is believed to be lognormal \citep{vazquezsemadeni94,passot.98,padoan.99,kritsuk...07}.
Some hints of power-law tails, however, have been noticed in numerical simulations of the 
self-gravitating turbulent interstellar medium 
\citep[ISM; e.g.,][]{scalo...98,klessen00,dib.05,slyz...05,vazquezsemadeni...08,federrath...08,collins...10}. More
recently similar tails were also found in high dynamic range observations of star-forming molecular 
clouds \citep{kainulainen...09,lombardi..10}. While it is understood that the density PDF holds the key to 
phenomenology of star formation \citep[e.g.,][]{padoan.02,krumholz.05,hennebelle.08,cho.10}, the origin 
of the power-law tail in self-gravitating supersonic turbulence still awaits a credible explanation. 
A related question pertains to the power index value for the tail. 

\citet{slyz...05} find a slope 
of $-1.5$ in non-magnetic kpc-scale interstellar turbulence simulations, while 
\citet{collins...10} measured $-1.6$ in isothermal
adaptive mesh refinement (AMR) MHD simulations of supersonic molecular cloud turbulence 
in a 10~pc box. Is there a universal power index value that applies to self-gravitating 
isothermal turbulence? What determines the slope? To address these questions, we analyze a deep 
AMR simulation with a linear dynamic range of $5\times10^5$ that follows the star formation 
process from turbulent initial conditions on a scale of a few pc down to a few AU. 

We describe the simulation detail
in the following section, while Section~3  presents the analysis of the density distribution and provides
testable predictions for the power index values. Section~4 discusses the limitations 
of the model and effects of the magnetic field on the density PDF. The final section outlines
conclusions.

\section{Numerical experiment}
Our star formation simulation was performed with the ENZO code for cosmology
and astrophysics \citep{osheaetal05} and discussed earlier in \citet{padoan...05}.
We solve the hydrodynamic equations, including self-gravity and a 
large-scale random force to drive the turbulence. We also adopt an 
isothermal equation of state and periodic boundary conditions. 
In this simulation, AMR is automatically 
carried out in collapsing regions in order to properly resolve the 
Jeans length \citep{truelove.....97}. We use root grid of
$512^3$ cells and five AMR levels with a refinement factor of four. The computational 
box has a size $L=5$~pc, and the gravitational collapse of dense protostellar 
cores is resolved down to the scale of 2~AU. The temperature is uniform, 
$T=10$~K, and the sound speed is constant, $c_{\rm s}=0.2$~km~s$^{-1}$. The mean 
density $n_0({\rm H}_2)=500$~cm$^{-3}$ and the rms flow velocity of $1.1$km~s$^{-1}$, 
typical of molecular clouds on scales of $\sim5$~pc, correspond to sonic Mach 
number $M_{\rm s}\approx6$. The free-fall time 
\begin{equation}
t_{\rm ff}\equiv\sqrt{\frac{3\pi}{32G\rho}}\approx1.6~{\rm Myr},
\end{equation}
the dynamical time
\begin{equation}
t_{\rm dyn}\equiv\frac{L}{2M_{\rm s}c_{\rm s}}\approx2.3~{\rm Myr},
\end{equation}
and the virial parameter for this model
\begin{equation}
\alpha\equiv\frac{5\sigma^2_{3D}R}{3GM}\approx0.25,
\end{equation}
correspond to a $3.4\times10^3$~$M_{\odot}$ molecular cloud prone to collapse on
its free-fall timescale. Indeed, the dendrogram analysis applied to a snapshot from a
larger $1024^3$ simulation \citep{kritsuk...07} resembling the initial conditions adopted 
here, indicated the presence of gravitationally bound objects on essentially all
scales within the computational domain \citep{rosolowsky...08}.

We began the simulation as a uniform grid turbulence model by stirring the gas in 
the computational domain for $4.8t_{\rm dyn}$ with a large-scale random force that includes 40\% 
dilatational and 60\% solenoidal power and then, at $t=0$, turned the forcing off to continue 
the simulation with AMR and self-gravity for about $0.29t_{\rm dyn}\approx0.43t_{\rm ff}$.

\section{Effects of Self-gravity}
\begin{figure}
\epsscale{1.2}
\plotone{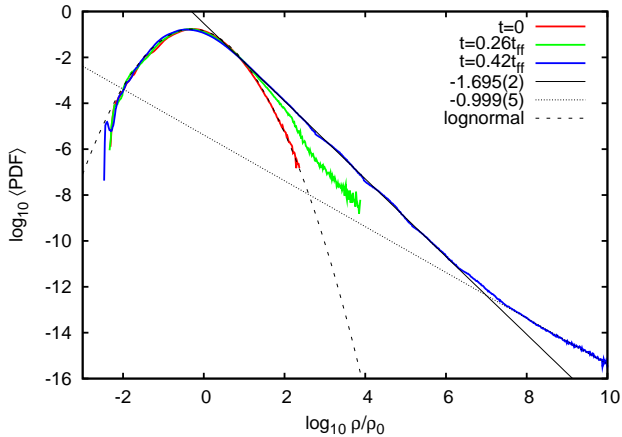}
\caption{Probability distribution functions for the mass density from an AMR simulation of
self-gravitating isothermal turbulence. The red line shows the initial conditions
corresponding to a driven, statistically steady Mach 6 turbulence with no self-gravity.
The green line shows the PDF for self-gravitating gas after $0.26t_{\rm ff}$ of evolution
from the initial conditions. The effective linear dynamic range of the simulation at 
this instance in time is 2048. The blue line shows a time-average PDF at $0.42\pm0.01t_{\rm ff}$;
time averaging helps to reduce the statistical noise at high densities. The dynamic range is
$5\times10^5$. The initial conditions can be approximated by a lognormal distribution (dashed
line). A power law with a slope of $-1.695\pm0.002$ (solid line) provides the best 
fit to the high-density tail at $\rho/\rho_0\in[10, 10^7]$. The break in the power index 
at $\rho/\rho_0\sim10^7$ marks a transition to rotationally supported cores (slope $-1$, 
dotted line).
}
\label{long}
\end{figure}

Figure~\ref{long} shows the density distributions in this simulation. The red line
corresponds to the initial condition at $t=0$, when we turn on self-gravity. The
density PDF at this time can be perfectly represented by a lognormal distribution
\citep{kritsuk...07}. Once gravity starts to operate, a power-law tail develops
at the high end of the distribution. After $0.26t_{\rm ff}$, when the creation of 
first-level AMR subgrids is triggered by the first collapsing objects, the density
distribution no longer remains lognormal. As the collapse of these first objects proceeds, 
followed by further grid refinement, an extended power-law tail emerges with a slope of about
$-1.7$. The tail departs from the initial lognormal distribution already 
at $\rho/\rho_0\sim10$ and continues straight for nearly 10~dex in probability and 
more than 6~dex in density. As
the simulation progresses, the slope continues to evolve slowly toward shallower values. 
The power index at the end of the simulation is $-1.67$. An even shallower tail at very 
high densities, $\rho/\rho_0>10^7$, develops as an indication of mass pile-up due to 
an additional support against gravity that comes from the conservation of angular 
momentum.\footnote{It may partly be also due to our enforced limit on the number of 
allowed refinement levels (five levels maximum), which eventually violates the 
\citet{truelove.....97} numerical stability condition.} The power index for this centrifugally
supported part of the density distribution is very close to $-1.0$. The fact that the
power law breaks at a density slightly in excess of $10^7\rho_0$ may indicate the minimum grid 
resolution requirement
for convergence in star formation simulations with sink particles (roughly $32,000$ for this
set of parameters), but the main focus of this Letter is on the origin of the extended power 
law at densities below $10^7\rho_0$. Why does it cover over six orders of magnitude in mass density
without a tiny bit of slope change? What fundamental physics is involved?

Let us first recall that supersonic turbulence is a multi-scale phenomenon shaping the structure 
of the mass distribution in molecular clouds \citep[e.g.,][]{vonweizs51,biglari.88,kritsuk..06}. 
At Mach numbers, $M_s>3$, the turbulence creates a ``fractal'' density distribution with the mass 
dimension of $D_{\rm m}\sim2.3$ \citep{elmegreen.96,kritsuk...07}. Since this mass dimension is
larger than the critical value for gravitational instability, $D_{\rm m}>D_{\rm crit}=2$, these 
highly inhomogeneous systems are still subject to gravitational collapse \citep{perdang90}.

\begin{figure*}
\epsscale{1.17}
\centerline{\plotone{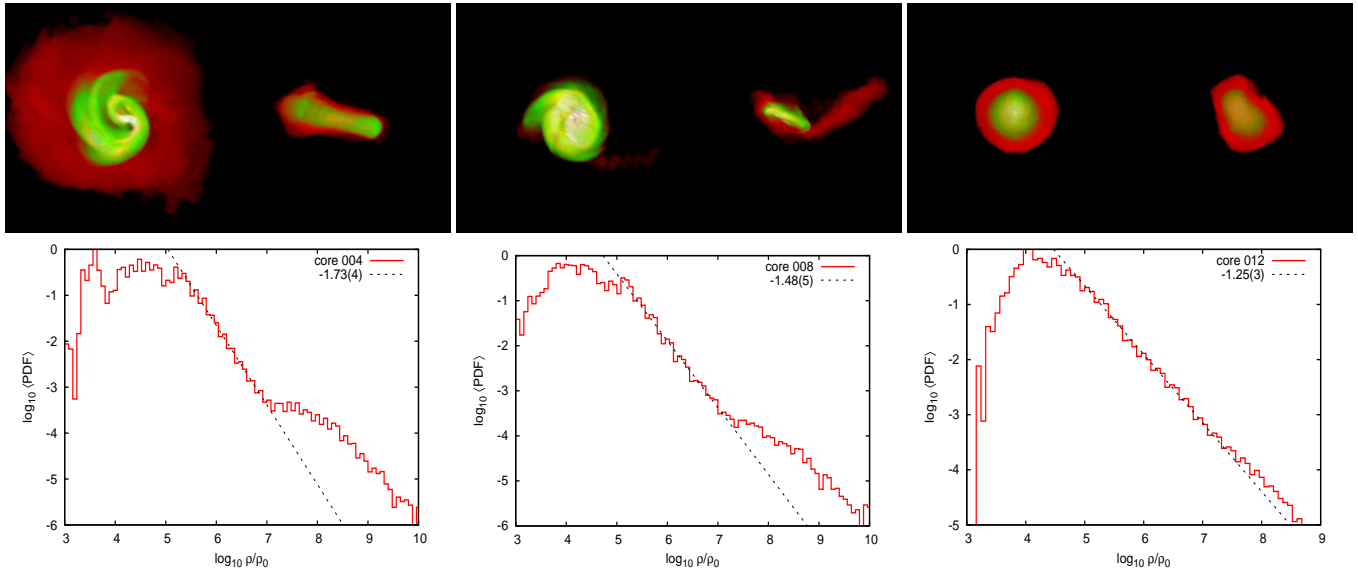}}
\caption{Volumetric rendering (top three panels) and PDFs (bottom) of the mass density 
for a sample of dense cores from the AMR simulation at $t=0.42t_{\rm ff}$. The data
represent $\sim350^3$ extractions at the highest grid resolution achieved in the simulation (2~AU).
The maximum densities are $\rho_{\rm max}/\rho_0=8.5\times10^{10}$, $3.2\times10^{10}$, and $6.0\times10^8$
for the cores from left to right, respectively.}
\label{extract}
\end{figure*}

The extent of the power law we obtain in the deep AMR simulation hints at the tail origin in 
the hierarchical nature of gravitational collapse of dense structures in molecular clouds. 
Let us take a look at the very bottom of the hierarchy, where AMR resolves collapsed protostellar 
cores. We used density fields for approximately cubic subvolumes centered on several selected 
dense cores to obtain the PDFs in the immediate vicinity of these objects. The linear size of 
these subvolumes is of order $0.005$~pc. Figure~\ref{extract} offers three example PDFs and 
volumetric rendering of the corresponding cores. Stretches of the power-law distributions are 
clearly present in all three cases, although the slopes vary from as shallow as $-1.25$ to as 
steep as $-1.75$.\footnote{Note that the rotation-induced pile-ups at highest densities are only 
visible in the two distributions that continue beyond $\rho/\rho_0=10^{10}$, while the third case 
shows only a hint of the pile-up at $\rho/\rho_0>10^7$. The first two cores have already developed 
relatively thin centrifugally supported disks, as can be seen in the renderings that show both 
face-on and edge-on views of the disks. The third core displays a rather modest flattening in the 
edge-on projection, indicating a weak rotation.} When the contributions from individual cores 
combine to form the density PDF for the whole computational domain, by some magic the resulting 
slope appears to be the same as that from the collapsing larger-scale structures characterized 
by a lower density, $\rho/\rho_0\in[10^2,10^5]$. 

The easiest way to solve this puzzle is to assume that a self-similar collapse solution 
would act as a strong attractor determining the form of the density PDF in hierarchical, 
turbulent, self-gravitating molecular clouds. There is a large inventory of (semi-)analytical 
solutions for the collapse of spherically symmetric isothermal configurations to choose from. 
\citet{whitworth.85} arranged these similarity solutions into a banded two-dimensional continuum 
embracing the limiting cases of {\em fast} \citep[][hereafter LP]{larson69,penston69} and {\em slow} 
\citep{shu77} collapse and their generalization by \citet{hunter77}. While this family of 
solutions describes gravitational condensation starting from a diverse set of idealized initial conditions, 
they have two important features in common. First, during the early stages of the evolution 
preceding the formation of a singularity at the center, all of them develop a $\rho\sim r^{-2}$ 
density profile.\footnote{The \citet{shu77} singular isothermal sphere (SIS) solution represents 
a hydrostatic $\rho\sim r^{-2}$ configuration from the outset.} Second, at late stages an 
expansion wave (EW) forms and propagates from the central singularity through the accreting 
material leaving behind a self-similar distribution with $\rho\sim r^{-3/2}$ at $r\rightarrow 0$.

It can be readily shown that the mass density PDF for a spherically symmetric configuration with a
$\rho=\rho_0(r/r_0)^{-n}$ density profile is a power law
\begin{equation}
dV=\frac{4}{3}\pi r_0^3\, d\left[\left(\frac{\rho}{\rho_0}\right)^{-3/n}\right]\propto d\left(\rho^m\right)
\label{32}
\end{equation}
with an index $m=-3/n$. Thus, formally, the similarity solutions generate power-law 
PDFs with a slope $m_{\rm LP}=-3/2$ corresponding to the $r^{-2}$ profile during the early collapse 
stages and with a combination of slopes $m_{\rm LP}=-3/2$ ($r^{-2}$ profile at lower densities) and 
$m_{\rm EW}=-2$ ($r^{-3/2}$ profile at higher densities) after the singularity has formed at the center. 
If the spherical symmetry is broken, for instance due to the presence of rotation, the situation becomes 
substantially more involved and hardly tractable analytically. \citet{whitworth...96} suggest that a
weak inward propagating compression wave that triggers the formation of the singularity would converge
incoherently on the center and interfere with any reflected EW. The interference would then cause a significant
degradation in the density profile at small radii. We believe this lowers expectations for the pure EW 
solution in realistic situations. However, the $r^{-2}$ density profile and 
the corresponding slope of the density PDF $m_{\rm LP}=-3/2$ could potentially be preserved if an
equilibrium singular disk with a flat rotation curve would form at the center 
\citep{norman..80,toomre82,hayashi..82}. Nevertheless, the disks formed in our simulation (Figure~\ref{extract})
typically show steeper density profiles ($\rho\sim r^{-3}$) and rotation curves that 
peak at $\sim10$~AU and then monotonically decline with radius up to $R\sim100$~AU. Higher
resolution would be required to follow the formation and structure of these disks with AMR 
properly.

The PDF slope $m=-1.67$ at the end of the simulation continues to evolve slowly toward shallower
values, but may still remain steeper than $-1.5$ corresponding to the $r^{-2}$ density profile. 
Neither does it approach $-2$ of the EW solution. Interestingly, a similar slope of $-1.64$ was 
independently obtained by \citet{collins...10} in a driven super-Alfv\'enic AMR MHD turbulence 
simulation with $M_{\rm s}=9$. This might indicate that some physics is missing in our discussion 
above. Indeed, both simulations model self-gravitating supersonic turbulent flows capable of 
creating the initial conditions for very dynamic collapses involving masses much in excess of 
the Bonnor--Ebert critical mass \citep{bonnor56,ebert55}. Such situations will be better 
approximated by pressure-free (PF) collapse solutions \citep{shu77}. The final stages of 
self-similar spherically symmetric PF collapse prior to the formation of central 
singularity are characterized by the $\rho\sim r^{-12/7}$ density profile \citep{penston69} 
which corresponds to the PDF power-law tail with $m_{\rm PF}=-7/4$. The free-fall collapse 
approximation, however, always breaks down near the center where the effects of pressure inevitably 
become important, so the slope of the high end of the PDF should still converge to $m_{\rm LP}=-3/2$. 
We indeed observe a slope change 
from $-1.7$ to $-1.5$ at $\rho/\rho_0\approx10^{6.2}$, see Figure~\ref{long}. Since the ratio
of gravitational-to-pressure forces in the isothermal PF collapse $J^*\propto r^{2/7}$
\citep{penston69}, it scales with the mass density as $\rho^{-1/6}$. This weak dependence 
of $J^*$ on the density  is consistent with the appearance of break in slope at very high
densities.

The projected density of an infinite sphere with the $\rho\sim r^{-n}$ density distribution,
\begin{equation}
\Sigma(R)=2\int^{\infty}_0\rho\left(\sqrt{R^2+x^2}\right)dx\propto R^{1-n},
\end{equation}
also has a power-law PDF,
\begin{equation}
dS\propto d \left(\Sigma^{-\frac{2}{n-1}}\right)\propto d \left(\Sigma^p\right),
\end{equation}
but with a slope $p=-2/(n-1)$. For the LP, PF, and EW similarity solutions, $p=-2$, $-2.8$, and $-4$, 
respectively.

\begin{figure}
\epsscale{1.2}
\plotone{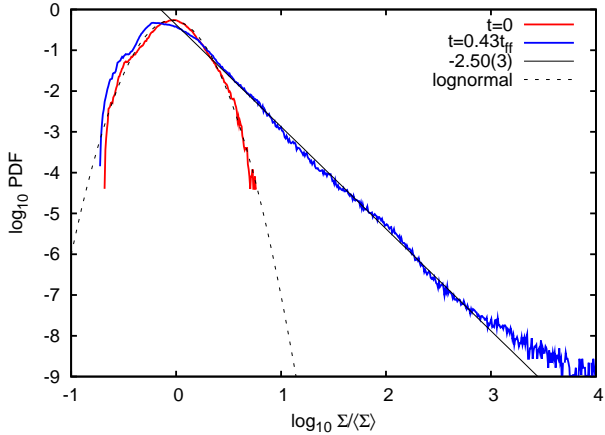}
\caption{Distributions of the projected gas density from the AMR
simulation at $t=0$ (red line) and at $t=043t_{\rm ff}$
(blue line). The initial distribution has a lognormal shape
(dashed line). The final distribution has an extended power-law
tail with a slope of $-2.50\pm0.03$ (solid line).}
\label{project}
\end{figure}

Figure~\ref{project} shows the projected mass density PDFs from the AMR simulation at $t=0$ (red line) and 
$t=0.43t_{\rm ff}$ (blue line). The blue line is based on a subset of the AMR data up to 8~AU 
in resolution; only the first four levels of fine mesh were used to make the plot. Similar to the three-dimensional density PDF, 
the initial distribution is well represented by a lognormal. The evolved self-gravitating configuration shows 
a clear power-law tail at high column densities with a slope of $-2.50\pm0.03$. The actual slope uncertainty 
may be larger than the formally determined value of $\pm0.03$. The distributions in 
Figure~\ref{project} show the average PDFs for all three projections, while only one projection would be
available in real observations. The root grid-based high-density tails for individual projections show some 
bumps (very similar to those seen in Figure~4 of \citet{kainulainen...09} and some straight sections with slopes 
broadly varying from as flat as $-2$ to steeper than $-3$. Overall, power index $p=-2.5$ measured in the 
simulation is right in between the predicted slopes for the PF and LP solutions, but the EW option ($p=-4$) 
seems to be rejected. We expect shallower slopes in simulations with lower Mach numbers or in longer turbulence 
decay simulations without continuous resupply of kinetic energy from random forcing. Finally, a hint of a 
shallower slope at $\Sigma>10^{2.8}$ may indicate the effect of centrifugal support, similar to the slope 
flattening in Figure~\ref{long}.

\section{Discussion}
We do not expect significant differences between driven and decaying turbulence 
models within the fraction of the first free-fall time in this AMR run \cite[see also][]{offner..08}.
By the end of the simulation, the system would still retain $>70$\% of the kinetic energy delivered
by the stirring force, if self-gravity were not included. However, self-gravity slightly 
increases the kinetic energy of a turbulent system \citep[e.g.,][]{slyz...05}, thus the lack 
of forcing does not really make a big difference. Since the free-fall time is shorter 
than the dynamical time that determines the energy decay timescale, the density PDF
established by $t=0.43t_{\rm ff}$ can be only weakly sensitive to the lack of forcing.

Since the virial parameter for this model is rather small, the role of self-gravity can be 
somewhat exaggerated. In situations with $\alpha\approx1$ one should expect a weaker effect.
With $\alpha$ close to unity, assuming the same Mach number and domain size, the power-law 
tail will be shallower (more closely resembling the LP case) in a more extended density range 
and will depart from the initial lognormal distribution at densities somewhat higher than 
$\rho/\rho_0=10$. While our adopted low value of the virial parameter creates a 
distribution similar to that of the Taurus molecular cloud, a higher $\alpha$ value would 
perhaps produce a distribution more similar to that of the Lupus~I cloud 
\citep[see Figure~2 in][]{kainulainen...09}. Overall, it seems that cloud-to-cloud
variations in the virial parameter $\alpha$ and in the age of the cloud in combination with 
projection effects can account for the full diversity of high-density tails in the observed 
star-forming clouds \citep[Figure~4 of][]{kainulainen...09}.

In this discussion on the density PDF, we so far ignored the effects of magnetic fields that are 
important for star formation. Our recent isothermal and multiphase MHD turbulence calculations, 
however, both show that variations in the level of magnetization
of interstellar clouds make little or no difference for the high-density end of the density PDF 
\citep[e.g.,][]{kritsuk..10}. In the absence of self-gravity, the high end of
the distribution preserves its lognormal shape. In super-Alfv\'enic turbulence, the magnetic 
field strength also shows a weak correlation with the gas density of the form $B\sim\rho^{1/2}$
\citep{padoan.99,collins...10}. The most recent Zeeman splitting data
for molecular cores also indicate a slope of $0.65\pm0.05$ \citep{crutcher....10}. A similar
relation can be inferred theoretically for dynamically collapsing magnetized protostellar cores
\citep[e.g.,][]{scott.80}.
Assuming that the correlation $B\sim\rho^{\gamma}$ exists, we predict a similar power-law 
tail in the PDF of the magnetic field strength with a slope from $-3\frac{1}{2}$ to $-2\frac{1}{4}$ 
for the range of $\gamma\in[1/2,2/3]$ and $m\in[-7/4,-3/2]$. A power index of $-2.7$ measured by 
\citet{collins...10} is consistent with $\gamma\approx0.6$ at $\rho/\rho_0<10^3$ and with their PDF slope 
$m=-1.64$.

\section{Conclusions}
We found an intriguing agreement between the probability distribution of molecular 
cloud densities in recent observations \citep{kainulainen...09} and in a deep AMR 
simulation of self-gravitating, supersonically turbulent molecular cloud
\citep{padoan...05}. In both cases, star-forming clouds display strong deviations 
from lognormal density distribution in the form of power-law tails at high density. 
In contrast, clouds with no active star formation display purely lognormal
distributions of density. 

We attribute the origin of the tails to the fundamental self-similar
properties of the $r^{-2}$ isothermal collapse and $r^{-12/7}$ pressure-free collapse 
laws, which control the density profiles of collapsing structures. This allows us to 
predict the power-law indices for the mass density ($m\in[-7/4,-3/2]$) and for the 
projected density ($p\in[-2.8,-2]$) depending on the physical conditions in the parent 
molecular cloud ($M_{\rm s}$, $\alpha$, etc.) that broadly agree with both 
observations and simulation results.

Our results may suggest a reconciliation of various attempts to build a phenomenological
theory of star formation and will contribute to the interpretation of numerical 
simulations in terms of the proposed phenomenologies \citep[e.g.,][]{schmidt...10,cho.10}.

\acknowledgements
This research was supported in part by a NASA ATP grant NNG056601G,
by NSF grants AST-0507768, AST-0607675, and AST-0908740, and by TeraGrid allocations
MCA98N020 and MCA07S014.
We utilized computing resources provided by the San Diego Supercomputer Center
(IBM-P690 {\em DataStar}, SUN Microsystems {\em Triton})
and by the National Institute for Computational Sciences (Cray-XT5 {\em Kraken}).

\end{document}